\documentclass[a4paper,11pt]{article}
\usepackage{pos}
\usepackage{bm}
\usepackage{bbm}

\title{Novel description of 
$P_c(4312)^+$, $P_c(4380)^+$, and $P_c(4457)^+$ with double triangle cusps
}
 \ShortTitle{Novel description of 
$P_c(4312)^+$, $P_c(4380)^+$, and $P_c(4457)^+$}

\author*{Satoshi X. Nakamura}

\affiliation{
University of Science and Technology of China, \\
Hefei 230026, People's Republic of China}

\affiliation{
State Key Laboratory of Particle Detection and Electronics
(IHEP-USTC),\\ 
Hefei 230036, People's Republic of China}

\emailAdd{satoshi@ustc.edu.cn}

\abstract{
We propose a novel scenario 
for the peak structures,
usually interpreted as 
hidden charm pentaquark ($P_c$) contributions,
 in the LHCb's
$\Lambda_b^0\to J/\psi p K^-$ data.
The key idea is to utilize 
leading or lower-order singularities from 
double triangle mechanisms.
The singularities cause anomalous threshold cusps,
which are significantly more singular than the ordinary ones,
at the $\Sigma_c^{(*)}\bar{D}^{(*)}$ threshold. 
We demonstrate that the double triangle amplitudes 
interfere with other common mechanisms to create 
peak structure that fit well 
the $P_c(4312)^+$, $P_c(4380)^+$, and $P_c(4457)^+$ peaks 
in the data.
This picture is completely different from commonly used ones such as
hadron molecules and compact pentaquarks. 
Meanwhile, 
$P_c(4440)^+$ is included in the proposed model as a resonance 
with width and strength significantly smaller than previously
estimated. 
The proposed model can (partly) explain the current data 
for other processes where $P_c^+$ signals are expected
such as:
no $P_c^+$ signals in
the GlueX $J/\psi$ photoproduction data;
a possible signal only from $P_c(4440)^+$ in
the LHCb's $\Lambda_b^0\to J/\psi p \pi^-$ data. 
}

\FullConference{%
  *** 10th International Workshop on Charm Physics (CHARM2020), ***\\
  *** 31 May - 4 June, 2021 ***\\
  *** Mexico City, Mexico - Online ***
}

\begin{document}
\maketitle

\section{Introduction}

The recent LHCb data on $\Lambda_b^0\to J/\psi pK^-$ revealed 
three resonance(like) structures~\cite{lhcb_pc}.
The peaks are considered to be contributions from
pentaquark states called $P_c(4312)^+$, $P_c(4440)^+$, and $P_c(4457)^+$.
Because the 
$P_c^+$ masses are slightly below the $\Sigma_c(2455)\bar{D}^{(*)}$
thresholds~\footnote{
We follow the hadron name notation of Ref.~\cite{pdg}. 
In addition, we often
denote $\Sigma_c(2455)^{+(++)}$, $\Sigma_c(2520)^{+(++)}$,
$\Lambda_c(2595)^+$, and $\Lambda_c(2625)^+$
by $\Sigma_c$, $\Sigma_c^*$, $\Lambda^*_c$, and $\Lambda^{**}_c$, 
respectively.
$\Lambda^{(*,**)}_c$ and $\Sigma_c^{(*)}$ are also collectively denoted by 
$Y_c$.
We often suppress charge indices.
We also denote a baryon ($B$) meson ($M$) pair with a spin-parity $J^P$
by $BM(J^P)$.
},
one would be tempted to interpret 
$P_c^+$'s as $\Sigma_c(2455)\bar{D}^{(*)}$ molecules (bound states).
Still, a compact pentaquark interpretation is also possible.
$P_c^+$'s are expected to appear also in different processes. 
The $J/\psi$ photoproduction off a nucleon seems a promising candidate. 
However, the GlueX experiment found no evidence~\cite{gluex}.
This may indicate either that a photon couple weakly with 
the $P_c^+$ states, or that 
the $P_c^+$ peaks in $\Lambda_b^0\to J/\psi pK^-$ are due to
kinematical effects
and do not appear in the photoproduction.

\begin{figure*}
\begin{center}
\includegraphics[width=1\textwidth]{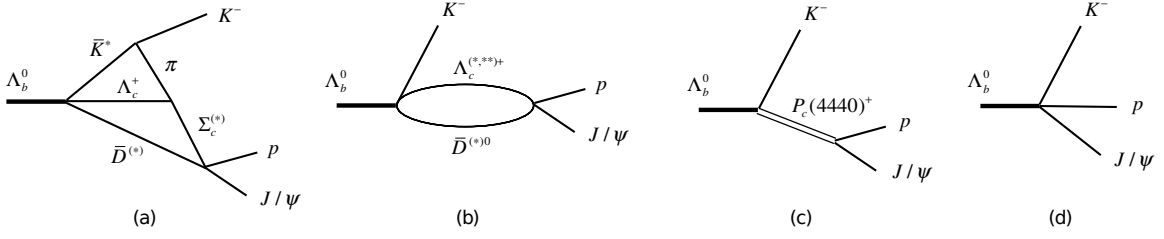}
\end{center}
 \caption{
$\Lambda_b^0\to J/\psi pK^-$ diagrams:
(a) double triangle;
(b) one-loop;
(c) $P_c(4440)^+$-excitation;
(d) direct decay.
Figures taken from Ref.~\cite{sxn_dts}. Copyright (2021) APS.
 }
\label{fig:diag}
\end{figure*}

In this work~\cite{sxn_dts}, we identify 
double triangle (DT) diagrams [Fig.~\ref{fig:diag}(a)]
some of which are kinematically allowed to occur at the classical level. 
Thus, the diagrams have
either the leading or lower-order
 kinematical singularities~\cite{s-matrix}.
The double triangle singularities (DTS) cause
anomalous threshold cusps.
We show that these cusps are
more singular than the ordinary threshold cusp.
Therefore, the DTS may be exploited to interpret resonancelike
structure. 
We demonstrate that the DT amplitudes
interfere with other common mechanisms of Figs.~\ref{fig:diag}(b) and \ref{fig:diag}(d)
to reproduce
the $P_c(4312)^+$, $P_c(4380)^+$, and $P_c(4457)^+$ peak structures
in the LHCb data.
Only one resonance is required to describe the $P_c(4440)^+$ peak.
We find that $P_c(4440)^+$ from our analysis has
width and strength significantly smaller than the LHCb's result.
This new interpretation of the $P_c$ signals is also (partly) consistent
with other data such as the $J/\psi$ photoproduction and 
$\Lambda_b^0\to J/\psi p \pi^-$ data~\cite{Pc_lhcb2} that seem to show 
only a $P_c(4440)^+$ signal.

\section{Model}
We consider mechanisms for $\Lambda_b^0\to J/\psi pK^-$
diagrammatically shown in Fig.~\ref{fig:diag}.
For loop diagrams [Fig.~\ref{fig:diag}(a,b)], 
the initial weak decays of
$\Lambda_b^0\to \Lambda^{(*,**)+}_c\bar{D}^{(*)}\bar{K}^{(*)}$
are assumed to be induced by color-favored quark mechanisms.
The $P_c(4440)^+$ amplitude [Fig.~\ref{fig:diag}(c)] is given 
in the Breit-Wigner form.
In each partial wave, 
a direct decay mechanism [Fig.~\ref{fig:diag}(d)] is included. 
We consider 
$J^P=1/2^-$, $3/2^-$, $1/2^+$, and $3/2^+$ 
partial waves;
$J^P$ denotes the spin-parity of $J/\psi p$.
Amplitude formulas can be found in Ref.~\cite{sxn_dts}.
The $Y_c\bar D^{(*)}$ pairs in the
DT and one-loop diagrams
are expected to be strongly interacting. 
We describe it with
a single-channel contact interaction model, and then combine it with
a perturbative transition to $J/\psi p$.
We assume to absorb 
other possible coupled-channel effects in complex
couplings fitted to the data. 

\section{Results}

\subsection{Singular behavior of double triangle amplitudes}

\begin{figure}[t]
\begin{center}
\includegraphics[width=1\textwidth]{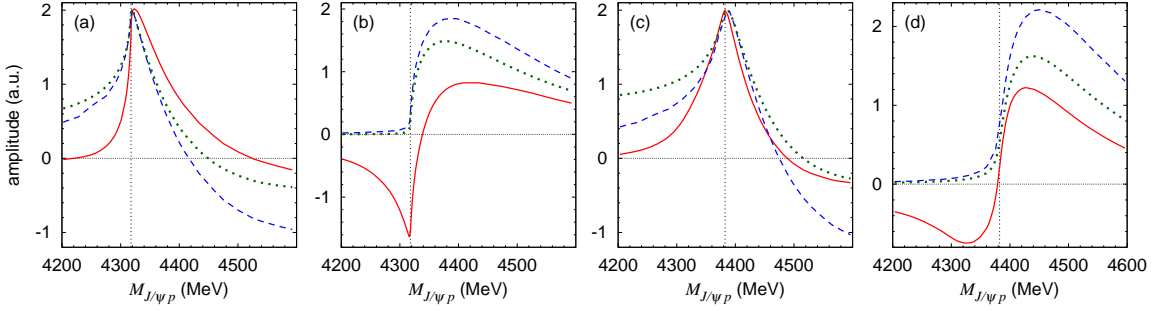}
\end{center}
 \caption{
Double triangle amplitudes.
(a) [(b)] The red solid curve shows
the real [imaginary] part of the double triangle amplitude
of Fig.~\ref{fig:diag}(a) with $\Sigma_c^{(*)}\bar D^{(*)}=\Sigma_c^{+}\bar D^{0}$;
$\Sigma^{(*)}_c\bar{D}^{(*)}\to J/\psi p$ is 
perturbatively considered.
By using $m_{\Lambda_c^+}=3$~GeV, 
they reduce to the blue dashed curves.
The $\Sigma_c^{+}\bar D^{0}$ one-loop amplitude is shown by
the green dotted curves.
All the amplitudes are normalized so that the real parts 
have the same peak height. 
The dotted vertical lines indicate
the $\Sigma_c^{+}\bar D^{0}$ thresholds.
(c) [(d)] 
The amplitudes shown are
obtained from those in
(a) [(b)] by replacing
$\Sigma_c^{+}$ with
 $\Sigma_c^{*+}$.
Figures taken from Ref.~\cite{sxn_dts}. Copyright (2021) APS.
 }
\label{fig:amp}
\end{figure}
The DT amplitude 
including $\Sigma_c^{+}\bar{D}^{0}\,(1/2^-)$ 
is shown by the red solid curves in Fig.~\ref{fig:amp}(a,b).
The amplitude is singular
near the $\Sigma_c^{+}\bar{D}^{0}$ threshold
because it has the leading singularity. 
A $\Sigma_c^{+}\bar{D}^{0}$
one-loop amplitude,
causing an ordinary threshold cusp,
is also shown by 
the green dotted curves.
The DT leading singularity
creates a more singular cusp. 
By setting 
the $\Lambda_c^+$ mass in the DT amplitude at
a hypothetically heavy value (3~GeV), 
the DT amplitude has 
only the
$\Sigma_c^{+}\bar{D}^{0}$ threshold singularity.
This is confirmed by the blue dashed curves
shown in the figure.
Similarly, we plot in Fig.~\ref{fig:amp}(c,d) 
amplitudes obtained from those in 
Fig.~\ref{fig:amp}(a,b) 
by replacing $\Sigma^{+}_c$ 
with $\Sigma^{*+}_c$.
This DT amplitude with $\Sigma^{*+}_c\bar{D}^0$ has 
the lower-order singularity and, thus, 
it is less singular than the 
leadingly singular
DT amplitude with $\Sigma^{+}_c\bar{D}^0$.
Still, both are 
more singular than the ordinary threshold cusp.

\begin{figure}[t]
\begin{center}
\includegraphics[width=.5\textwidth]{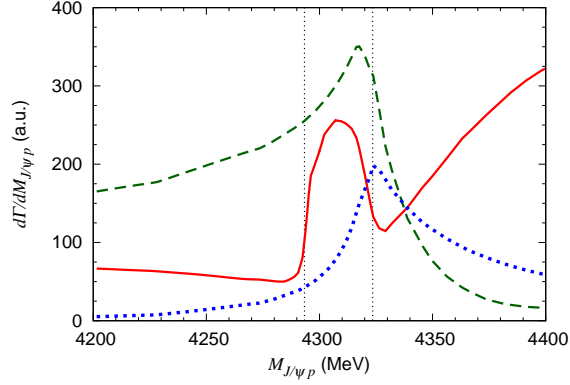}
\end{center}
 \caption{
Formation of the $P_c(4312)^+$ peak from interference of different amplitudes. 
The blue dotted curve is the differential decay width
($d\Gamma/dM_{J/\psi p}$) solely from
the double triangle amplitude including the $\Sigma_c\bar{D}$ pair.
By adding a direct decay amplitude coherently, the green dashed curve is
 obtained. 
The red solid curve is obtained
by further adding the $\Lambda_c^+\bar{D}^{*0}$ one-loop diagram.
$Y_c\bar{D}^{(*)}\to J/\psi p$ is 
perturbatively considered.
The dotted vertical lines indicate thresholds for,
 from left to right, $\Lambda_c^+\bar{D}^{*0}$ and
 $\Sigma_c(2455)^{++}D^-$, respectively.
 }
\label{fig:pc}
\end{figure}
How is a $P_c$ peak created from a DT amplitude ? 
The DT amplitude including $\Sigma_c\bar{D}$ alone gives
the $M_{J/\psi p}$ distribution
shown by the blue dotted curve in 
Fig.~\ref{fig:pc}.
A peak in the spectrum is located at the $\Sigma_c\bar{D}$ threshold,
as expected from Fig.~\ref{fig:amp}.
This peak position does not agree with $P_c(4312)$.
By including the direct decay amplitude coherently, 
we obtain the green dashed curve with a peak slightly below the
threshold. 
By further including the $\Lambda_c^+\bar{D}^{*0}$ one-loop amplitude, 
the red solid curve is obtained; now the peak position agrees with $P_c(4312)$.

\subsection{Analyzing the LHCb data}

\begin{figure*}[t]
\begin{center}
\includegraphics[width=1\textwidth]{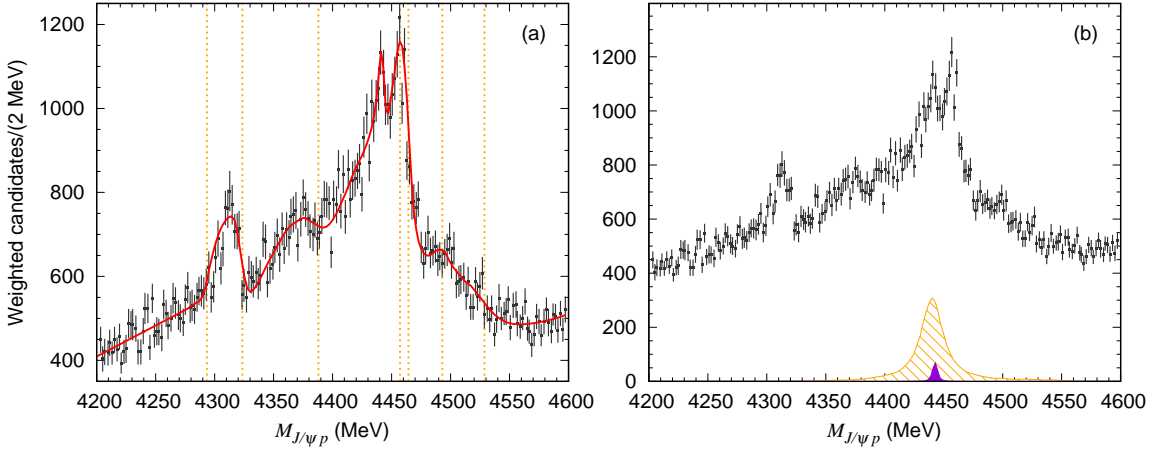}
\end{center}
 \caption{
(a) Comparison with the LHCb data
($\cos\theta_{P_c}$-weighted samples)~\cite{lhcb_pc}
for $J/\psi p$ invariant mass $(M_{J/\psi p})$ distribution of
$\Lambda_b^0\to J/\psi pK^-$.
The red solid curve is from our model.
The dotted vertical lines indicate thresholds for,
 from left to right, $\Lambda_c^+\bar{D}^{*0}$, $\Sigma_c(2455)^{++}D^-$,
$\Sigma_c(2520)^{++}D^-$, 
$\Lambda_c(2595)^+\bar{D}^0$,
$\Sigma_c(2455)^{++}D^{*-}$, 
$\Lambda_c(2625)^+\bar{D}^0$,
and $\Sigma_c(2520)^{++}D^{*-}$, respectively.
(b)~$P_c(4440)^+$ contribution.
The orange striped peak is the $P_c(4440)^+$ contribution from the LHCb analysis.
The solid violet peak is the $P_c(4440)^+(3/2^-)$ contribution from our
 model; the interference excluded.
Figure is (partly) taken from Ref.~\cite{sxn_dts}. Copyright (2021) APS.
 }
\label{fig:comp-data}
\end{figure*}
We consider in our 
$\Lambda_b^0\to J/\psi pK^-$ model:
(i) DT mechanisms with
$\Sigma_c(2455)\bar{D}\, (1/2^-)$,
$\Sigma_c(2520)\bar{D}\, (3/2^-)$,
$\Sigma_c(2455)\bar{D}^*\, (1/2^-)$,
$\Sigma_c(2455)\bar{D}^*$ $(3/2^-)$,
$\Sigma_c(2520)\bar{D}^*\, (1/2^-)$, and
$\Sigma_c(2520)\bar{D}^*\, (3/2^-)$;
(ii) one-loop mechanisms with
$\Lambda_c^+\bar{D}^{*0}\, (1/2^-)$,
$\Lambda_c(2595)^+\bar{D}^{0}\, (1/2^+)$, and
$\Lambda_c(2625)^+\bar{D}^{0}\, (3/2^+)$;
(iii) $P_c(4440)^+$ mechanism;
(iv) direct decay mechanisms.
Fitting parameters come from 
each mechanism in the items (i)-(iii) with an adjustable complex overall factor;
$2\times 10$ parameters.
Four parameters from 
direct decay mechanisms (iv) each of which has a real coupling.
Two parameters from the $P_c(4440)^+$ mass and width.
One parameter from a repulsive $\Lambda_c^+\bar{D}^{*0}$ interaction strength. 
Because the full amplitude has an arbitrariness of 
the overall absolute normalization, 
we have 26 parameters in total. 

Each of $Y_c\bar D^{(*)}(J^P)$ interactions are examined if the fit 
prefers an attraction or repulsion.
We found attractions for
$\Sigma_c(2455)\bar{D}\, (1/2^-)$,
$\Sigma_c(2520)\bar{D}\, (3/2^-)$,
$\Sigma_c(2455)\bar{D}^*\, (1/2^-)$,
$\Sigma_c(2455)\bar{D}^*\, (3/2^-)$,
$\Lambda_c(2595)^+\bar{D}^{0}$ $(1/2^+)$,
$\Lambda_c(2625)^+\bar{D}^{0}$ $(3/2^+)$,
and repulsions for
$\Sigma_c(2520)\bar{D}^*\, (1/2^-)$,
$\Sigma_c(2520)\bar{D}^*\, (3/2^-)$,
$\Lambda_c^+\bar{D}^{*0}\, (1/2^-)$.
Then we use a fixed coupling 
for the attraction
so that
the scattering length is $a \sim 0.5$~fm~\footnote{
The scattering length $(a)$ is related to the phase shift $(\delta)$ by 
$p\cot\delta=1/a + {\cal O}(p^2)$.} 
We fit
the repulsive $\Lambda_c^+\bar{D}^{*0}\,(1/2^-)$ coupling
to the data to keep a good fit quality;
$a \sim -0.4$, $-0.2$, and $-0.05$~fm
for $\Lambda\sim 0.8$, $1$, and $1.5$--$2$~GeV, respectively 
($\Lambda$: common cutoff in the form factors).
The same coupling strength is used for
the other repulsive $Y_c\bar D^{(*)}(J^P)$ channels.
Spectrum peak positions are not very sensitive to the $a$ values
since they are essentially determined by the kinematical effects. 
We use $\Lambda=1$~GeV at each interaction vertex;
the result does not significantly depend on the cutoff value.
An exception is applied to
the direct decay amplitudes for which 
different cutoffs on $p_{\bar{K}}$ are used
so that their $M_{J/\psi p}$ distribution is similar to the phase-space shape.

We compare the calculation, after smearing with the experimental resolution,
with the LHCb data~\cite{lhcb_pc}
in Fig.~\ref{fig:comp-data}(a).
The data are well fitted by
our full model shown by the red solid curve.
The considered mechanisms cause the kinematical effects to describe well 
the $P_c(4312)^+$, $P_c(4380)^+$, and $P_c(4457)^+$ 
peak structures.
We utilized a pole ($J^P=3/2^-$ in the figure) to fit 
only the $P_c(4440)^+$ peak.
We varied the cutoff over $\Lambda=$ 0.8--2~GeV and change 
$J^P$ for $P_c(4440)^+$ over $J^P=1/2^\pm$ and $3/2^\pm$;
the quality of the fit 
does not change significantly.

In Fig.~\ref{fig:comp-data}(b),
we show 
the $P_c(4440)^+(3/2^-)$ contribution from our analysis by 
the violet solid peak;
an interference contribution is excluded.
This contribution is described by the Breit-Wigner
mass and width of
4443.1$\pm 1.4$~MeV and 2.7$\pm 2.4$~MeV,
respectively.
These values can be 
compared with those from the LHCb analysis~\cite{lhcb_pc},
$4440.3\pm 1.3^{+4.1}_{-4.7}$~MeV and 
$20.6\pm 4.9^{+8.7}_{-10.1}$~MeV,
shown by the orange striped peak in 
Fig.~\ref{fig:comp-data}(b).
The width from our analysis is significantly narrower. 
Also, our $P_c(4440)^+$ contribution 
is smaller by a factor of $\sim$22
than the LHCb's estimate:
${\cal R}\equiv {\cal B}(\Lambda_b^0\to P_c^+K^-){\cal B}(P_c^+\to J/\psi p)
/{\cal B}(\Lambda_b^0\to J/\psi pK^-)=1.11\pm 0.33^{+0.22}_{-0.10}$~\%.
This large difference comes from different fitting strategies.
The LHCb used incoherent $P_c(4440)^+$ and $P_c(4457)^+$ contributions
to fit the large structure at $M_{J/\psi p}\sim 4450$~MeV.
On the other hand,
we utilized the kinematical effects to
describe a large portion of the structure,
and fit the remaining small spike with
the $P_c(4440)^+$ and its interference. 

The LHCb found an evidence for $P_c^+$ 
also in $\Lambda_b^0\to J/\psi p \pi^-$.
In their $M_{J/\psi p}$ distribution [Fig.~3(b) of \cite{Pc_lhcb2}],
the $M_{J/\psi p}$ bin of $P_c(4440)^+$ seems to be  enhanced.
However, no visible enhancement is found for 
the other $P_c^+$'s.
This observation is actually consistent with our model's expectation because
$\Lambda_b^0\to J/\psi p \pi^-$ does not have a relevant DT mechanism
but can share the $P_c(4440)^+$ excitation mechanism.
However, the 
$\Lambda_b^0\to J/\psi p \pi^-$ data
may conflict with some other $P_c^+$ models.
The $P_c^+$ signals in $\Lambda_b^0\to J/\psi p \pi^-$
are still inconclusive due to 
the limited quality of the data.
Forthcoming LHCb Run II data on $\Lambda_b^0\to J/\psi p \pi^-$
might seriously challenge the models.

\section{Summary}

We developed a model for $\Lambda_b^0\to J/\psi p K^-$ 
and analyzed the $M_{J/\psi p}$ distribution data from
the LHCb.
The double triangle cusps and their interference with the common
mechanisms describe well 
the $P_c^+$ structures.
$P_c(4440)^+$ is the only resonance in our analysis, and 
its width and strength are much smaller than those from the LHCb analysis.
This interpretation of the $P_c^+$ peaks is completely different 
from the commonly used hadron molecules
and compact pentaquarks. 
The DT cusps can appear in different processes and thus they 
should now be an option to interpret 
resonancelike structures near thresholds.

\begin{acknowledgments}
This work is in part supported by 
National Natural Science Foundation of China (NSFC) under contracts 
U2032103 and 11625523, 
and also by
National Key Research and Development Program of China under Contracts 2020YFA0406400.
\end{acknowledgments}

\end{document}